\documentstyle[aps,prb,multicol,graphicx]{revtex}
\def\Tr{\mbox{Tr}}
\def\D{\mbox{D}}

\def\max{\mbox{max}}
\def\S{\mbox{S}}

\def\I{\mbox{I}}
\def\M{\mbox{M}}
\def\B{\mbox{B}}

\title {Shot Noise at High Temperatures}
\author{D.B. Gutman and Yuval Gefen }
\address{
Department of Condensed Matter Physics, Weizmann Institute of
Science 76100 Rehovot, Israel}
\begin{document}
\date{\today}
\maketitle
\begin{abstract}
We consider the possibility of measuring non-equilibrium
properties of the current correlation functions at high
temperatures (and small bias). Through  the example of the third
cumulant of the current  (${\cal{S}}_3$) we demonstrate that odd
order correlation functions represent  non-equilibrium physics
even at small external bias and high temperatures. We calculate
${\cal{S}}_3=y(eV/T) e^2 I$ for a quasi-one-dimensional diffusive
constriction. We calculate the scaling function $y$ in two
regimes: when the scattering processes are purely elastic and when
the inelastic electron-electron scattering is strong. In both
cases we find that $y$ interpolates between  two constants. In the
low (high) temperature limit $y$ is strongly (weakly) enhanced
(suppressed) by the electron-electron scattering.
\end{abstract}
\begin{multicols}{2}
\section{introduction}
\label{sec:introduction} Since its discovery by Schottky in 1918,
noise in electrical circuits has been thoroughly investigated.
Numerous studies over the past decade, both experimental
\cite{experiments} and theoretical (see Refs.
\cite{Kogan_book,Blanter} for reviews), emphasized the quantum and
the mesoscopic aspects of noise, addressing quite extensively the
issue of non-equilibrium "shot" noise. Shot noise has been
employed for probing fundamental physics of interacting electrons
in the FQHE regime \cite{de-Piccioto}. Most of previous work
focused on current-current correlations, i.e., fluctuations
derived from second order current cumulants\cite{Gavish}
\end{multicols}
\begin{eqnarray}
\label{second} \S_2(x,x',\omega)=\frac{1}{2}\int
d(t-t')e^{-i\omega(t-t')}\langle\hat{I}(t,x)\hat{I}(t',x')+\hat{I}(t',x')\hat{I}(t,x)\rangle
,
\end{eqnarray}
\begin{multicols}{2}
where $\hat{I}(t,x)$  denotes the instantaneous current operator
integrated trough the entire cross-section (at point $x$); $x$ is
the direction of the flow  of the d.c. current. Equilibrium noise
is governed by the  Callen-Welton relation, also known as
Fluctuation Dissipation Theorem (FDT). It connects the Ohmic part
of the  conductance $G(\omega)$ with the equilibrium current noise
(${\cal{S}}_2=\S_2-I^2$ , $I \equiv \langle\hat{I}\rangle$)
\begin{eqnarray}
\label{eqnoise} {\cal{S}}_2^{eq}(\omega)=\hbar\omega
G(\omega)\coth\left(\frac{\hbar \omega}{2T}\right)\,\, .
\end{eqnarray}

Let us consider  a  two-terminal constriction subject to an
applied bias,  $eV$,  which is equal to the difference between the
chemical potentials of the left and the right reservoirs
$eV=\mu^L-\mu^R$. The constriction  length $L$ is much larger than
an elastic mean free path $l=v_f\tau$ ($L \gg l$). The value of
the d.c. electron current through the system is given by Ohm's
law, $I=GV$. In the absence of electron-electron collisions the
second moment of the current fluctuations can be expressed in
terms of the equilibrium correlation function, eq.~(\ref{eqnoise})
(cf. Ref.~\cite{ALY,FDT_resembles})
\begin{eqnarray}&&
\label{cc1} {\cal{S}}_2(\hbar\omega)\!=\!\frac{1}{6}\bigg[
4{\cal{S}}^{eq}_2(\hbar\omega)\!+{\cal{S}}^{eq}_2(\hbar\omega\!+\!eV)+{\cal{S}}^{eq}_2(\hbar\omega\!-\!eV)
\bigg].
\end{eqnarray}
Depending on the  ratio between the temperature of the system,
$T$, the frequency under discussion, $\omega$,  and the applied
voltage,  $V$,  one may distinguish among different regimes. In
the limit $( eV \gg \max\{kT,\omega\})$ the noise, eq.(\ref{cc1}),
is essentially non-equilibrium (shot noise), and is proportional
to the absolute value of the d.c.  electric current. For $eV \ll
\max\{kT,\omega\}$   ${\cal{S}}_2$ is mostly  thermal, or Nyquist
noise. These observations are not system dependent, but  rather
follow from general considerations, namely  the properties of
various  operators under  time reversal transformation.

Since the current operator changes sign under time reversal
transformation, any even-order correlation function of the current
fluctuations (e.g. ${\cal{S}}_2$) taken at zero frequency is
invariant under this operation. Assuming that current correlators
are functions of the average current, $I$, it follows that
even-order   correlation functions depend only on the absolute
value of the electric current (and are independent of the
direction of the current). In the Ohmic regime this  means that
even-order current correlation functions (at zero frequency)  are
even functions of the applied voltage. Evidently, this general
observation  agrees with the result eq.~(\ref{cc1}) for the
second-order correlation function ${\cal{S}}_2$ in the diffusive
junction.

By contrast,  odd-order current correlation functions change their
sign under time reversal transformation. In other words, such
correlation functions depend on the direction of the current, and
not only on its absolute value. Therefore in  the Ohmic regime,
odd-order correlation functions of current are odd-order functions
of the applied voltage. This condition automatically guarantees
that odd-order correlation functions {\it vanish}  at thermal
equilibrium.

 The present analysis focuses on the simplest, yet
non-trivial, example  of manifestly  out-of-equilibrium
correlator, namely the third order {\it current correlation
function}  $\S_3$. This is the lowest order correlation function
which is dominated by non-equilibrium fluctuations even in the
regime where the applied bias is small compared with temperature.
There are two situations where the study of shot noise under such
conditions is called for. Firstly, when the lowering of the
temperature of the electron is not facilitated. Secondly, and more
interestingly, when one is interested in exploring the
correlations in the system above a certain  critical (or a
characteristic) temperature. On one hand this requires to keep the
temperature relatively high. On the other hand, to study low
energy features, one is restricted to low  values of the voltage.

In this work we argue and subsequently show that for the third
order (and in fact all odd order) {\it current cumulant},
out-of-equilibrium fluctuations are not masked by large
equilibrium noise: ${\cal{S}}_3$  remains linear in voltage (and
{\it  temperature independent})  at high temperatures. The first
part of our analysis (Section\ref{sec:elastic}) focuses on  the
temperature dependence of ${\cal{S}}_3$,  in the limit where  the
inelastic length is much larger  than the system's size. We find
that ${\cal{S}}_3$ is proportional to the the d.c. current at all
temperatures, and calculate how the proportionality coefficient
interpolates between the two asymptotic values, $1/15$ and $1/3$,
for $T \ll eV$ and $T\gg eV$ respectively. In the second part of
this paper (Section \ref{sec:inelastic}) we study the effects of
{\it inelastic} electron-electron scattering. We specifically
consider the limit of $l_{\rm in}/L \ll 1$, where $l_{\rm in}$ is
the inelastic electron-electron mean free path and $L$ is the
length of the conductor. We find that as compared with the elastic
case ${\cal{S}}_3$ is weakly suppressed in the high temperature
limit (by $\sim 15 \%$)but is highly enhanced in the low
temperature limit (cf. Table \ref{table1}). Similarly to the
elastic case ${\cal{S}}_3$ is not masked by thermal fluctuations
in the high temperature limit.

Our analysis employs the recently developed non-linear
$\sigma$-model--Keldysh technique \cite{KA,Nayak,GG}. This
approach has been applied to study non-equilibrium noise in the
presence of disorder and electron-electron interactions \cite{GG}.
The main steps to be followed in the analysis below include
finding the saddle point of the effective action (eqs. \ref{e25}
and \ref{ansatz} for the elastic and the inelastic cases
respectively) and expanding in soft modes around them (eqs.
\ref{corr1} and \ref{inelasticd}). The saddle point (or the
approximate one in the inelastic case) are connected with the
single-particle distribution function, while the soft modes
describe the dynamics of the density fluctuations. Both are
significantly modified by strong inelastic scattering.

There are of course other methods that can be employed to study
non-equilibrium fluctuations. One appealing candidate would be the
Shulman-Kogan \cite{Shulman} version of the Boltzmann-Langevin
scheme. This method work successfully when ${\cal{S}}_2$ is
studied.  In Section \ref{sec:kinetic} we present a short
discussion which shows, arguably in an unexpected way, that the
traditional kinetic equation approach  breaks down when higher
order cumulants are addressed. Section \ref{sec:discussion}
includes a brief discussion.

\section{The Elastic Case}
\label{sec:elastic}
 As has been shown in
Ref.\cite{Quantum_Galvanometr}, by modeling the measurement
procedure as coupling of the system to a ``quantum galvanometer'',
the emerging  mathematical object to be studied is
\begin{eqnarray}
\label{s3} \S_3(x,t;x',t';x'',t'')=\langle T_c
\hat{I}_2(x,t)\hat{I}_2(x',t')\hat{I}_2(x'',t'') \rangle \,\, .
\end{eqnarray}
Here $x$ is a coordinate measured along a quasi one-dimensional
wire ($0 \le x\le L$) of cross-section ${\cal{A}}$;  $T_c$ is the
time ordering operator along the Keldysh contour and $I_2$ is a
quantum component of the current. Since the process we consider is
stationary, the correlation function depends only on  differences
of time, $\S_3(t-t';t'-t'')$. In  Fourier space it can be
represented as
\end{multicols}
\begin{eqnarray}
\S_3(\omega_1,\omega_2)=\int_c d(t-t') d(t'-t'')
e^{-i\omega_1(t-t')-i\omega_2(t'-t'')}\S_3(t-t',t'-t'')\,\, .
\end{eqnarray}
\begin{multicols}{2}
(Here we have assumed that the frequencies $\omega_1,\omega_2$ are
small compared with the inverse diffusion time along the wire,
such that the current fluctuations are independent of the spatial
coordinate).  We next evaluate the expression, eq.~(\ref{s3}), for
a particular system, in the hope that the qualitative properties
we are after are  not strongly system dependent. In the present
section we consider non-interacting electrons in the presence of a
short-range, delta correlated and weak disorder potential
($\epsilon_f \tau \gg \hbar$, where $\tau$ is the elastic mean
free time and $\epsilon_f$ is the  Fermi energy). To calculate
$\S_3$ we employ the $\sigma$-model formalism, recently put
forward for dealing with non-equilibrium diffusive systems (for
details see Ref.\cite{GG}). As was stressed in our previous work
this approach is a generalization of the Boltzmann-Langevin
kinetic approach \cite{Shulman}. It is comparable to the latter as
long as the kinetic  approach is applicable to diffusive systems.
A direct employment of the kinetic approach, assuming the random
force term to be short-ranged correlated in space, does not give
correctly the the higher-than-two cumulants of the current. The
reasons for that are discussed below (cf. Section
\ref{sec:kinetic}).

The Hamiltonian of our system of non-interacting electrons in
disordered systems is:
\begin{eqnarray}
\label{Hamiltonian} H_0=\int_{\rm Volume} d{\bf r} \bar{\Psi}({\bf
r})\bigg[-\frac{\hbar^2}{2m}(\nabla-{\bf a})^2 + U_{\rm
dis}\bigg]\Psi({\bf r}).
\end{eqnarray}
Here $c{\bf a}/e$ is a vector potential. The disorder potential is
$\delta$-correlated:
\begin{equation}
\label{<>} \langle U_{\rm dis}({\bf r})~U_{\rm dis}({\bf
r}')\rangle = \frac{1}{2\pi \nu \tau} ~ \delta({\bf r} - {\bf
r}'),
\end{equation}
where $\nu$ is the  density of states at the Fermi energy.

Following the procedure outlined in Ref. \cite{GG}, employing the
diffusive approximation and focusing on the out-of-equilibrium
system,  one can write a generating functional, expressed  as a
path integral over a bosonic matrix field $Q$
\begin{eqnarray}&&
\label{gen1} Z[{\bf a}] =\int {\cal D}Q \exp(iS[Q,{\bf a}])\,\, .
\end{eqnarray}
\special{push color Blue} Here the integration is performed  over
the manifold
\begin{eqnarray}&&
\label{manifold} \int Q(x,t,t_1)Q(x,t_1,t')dt_1=\delta(t-t'),
\end{eqnarray}
the effective action is given by \special{pop color}
\begin{eqnarray}&&
\label{action1} iS[Q,{\bf a}]=-\frac{\pi\hbar\nu}{4}\Tr\big\{D
\left( \nabla Q+i[{\bf a}_\alpha \gamma^\alpha,Q] \right)^2 +4i
\hat{\epsilon} Q\big\}\,\,,
\end{eqnarray}

and
\begin{eqnarray}&&
\gamma_1=\left(\matrix {1 & 0\cr 0 & 1 \cr}\right) , \;
\gamma_2=\left(\matrix {0 & 1\cr 1 & 0 \cr}\right) \, .
\end{eqnarray}
$\Tr$ represents summation over all spatio-temporal and Keldysh
components. Here ${\bf a}_1$ and ${\bf a}_2$ are the Keldysh
rotated classical  and quantum components of ${\bf a}$. Hereafter
we focus our attention on ${\bf a}_1$, ${\bf a}_2$, the components
in the direction along the wire. The third order current
correlator may now be expressed as functional differentiation of
the generating functional $Z[a]$ with respect to $a_2$
\begin{eqnarray}
\label{z27}
\!\S_3(\!t_1\!-\!t_2,t_2\!-\!t_3\!)\!=\frac{ie^3}{8}\frac{\delta^3Z[a]}{\delta\!a_2(x_1,t_1)\delta\!a_2(x_2,t_2)\delta\!a_2(x_3,t_3)}.
\end{eqnarray}
Performing this  functional differentiation  one obtains the
following result
\begin{eqnarray}&&
\label{z28} \S_3(\!t_1\!-\!t_2,t_2\!-\!t_3\!)\!
=\frac{e^3{\cal{A}}(\pi\hbar\nu D)^2}{8}\nonumber \\&&
\bigg\langle\frac{1}{2}\hat{\M}(x_1,t_1)\hat{\I}^D
(x_2,x_3,t_2,t_3)+\nonumber \\&& (x_1,t_1\leftrightarrow\!x_3,t_3)
+ (x_1,t_1\leftrightarrow\!x_2,t_2)+ \nonumber \\&&
\frac{\pi\hbar\nu
D}{8}\hat{\M}(x_1,t_1)\hat{\M}(x_2,t_2)\hat{\M}(x_3,t_3)\bigg\rangle
\,\,  .
\end{eqnarray}
Here we have defined
\begin{eqnarray}&&
\label{e33}
\hat{I}^D(x,x',t,t')=\Tr^K\!\bigg\{\!Q_{x,t,t'}\gamma_2Q_{x',t',t}\gamma_2-
\delta_{t,t'}\gamma_1\bigg\}\delta_{x,x'} ,
\end{eqnarray}
\begin{eqnarray}&&
\label{M}
\hat{\M}(x,t)=\Tr^K\!\bigg\{\!\int\!dt_1\bigg([Q_{x,t,t_1};\nabla
]~ Q_{x,t_1,t} \bigg)\gamma_2\bigg\}.
\end{eqnarray}
We employ the notations $Q(x,t,t')\equiv Q(x,t,t')$; $\Tr^K$ is
the trace  taken with respect to the Keldysh indices; $\langle
\rangle$ denotes a quantum-mechanical  expectation value. The
matrix $Q$ can be parameterized as
\begin{eqnarray}&&
\label{e24} Q=\Lambda\exp\left(W\right) \,\, ,
\end{eqnarray}
where
\begin{eqnarray}&&
\Lambda W+W \Lambda=0 \,\, ,
\end{eqnarray}
and $\Lambda$ is the saddle point of the action (\ref{action1})
\begin{eqnarray}&&
\label{e25} \Lambda(x,\epsilon)=\left(\matrix{1 & 2F(x,\epsilon)
\cr 0 & -1\cr}\right)\,\, .
\end{eqnarray}
The function $F$ is related to the single particle distribution
function $f$ through
\begin{eqnarray}&&
\label{e31} F(x,\epsilon)=1-2f(x,\epsilon) \,\, .
\end{eqnarray}
The matrix $W_{x,\epsilon,\epsilon'}$,  in turn, is parameterized
as follows:
\begin{eqnarray}
\label{c20} W_{x,\epsilon,\epsilon'}\!= \!\left( \matrix
{F_{x,\epsilon} \bar{w}_{x,\epsilon,\epsilon'} &
-w_{x,\epsilon,\epsilon'} +
F_{x,\epsilon}\bar{w}_{x,\epsilon,\epsilon'}F_{x,\epsilon'}\cr-\bar{w}_{x,\epsilon,\epsilon'}
& - \bar{w}_{x,\epsilon,\epsilon'}F_{x,\epsilon'}\cr} \right)  .
\end{eqnarray}
It is convenient to introduce the diffusion propagator
\begin{eqnarray}&&
(-i\omega
+D\nabla^2)\D(x,x'\omega)=\frac{1}{\pi\hbar\nu}\delta(x-x') \, \,
.
\end{eqnarray}
The absence of diffusive motion   in  clean metallic leads implies
that the  diffusion propagator must vanish at the end points of
the constriction. In addition,  there is  no current flowing in
the transversal direction (hard wall boundary conditions). It
follows that the component of the gradient of the diffusion
propagator in that direction (calculated at the hard wall edges)
must vanish as well. The correlation functions of the fields $w,
\bar{w}$ are then given by:
\end{multicols}
\begin{eqnarray}&&
\label{corr1} \langle w(x,\epsilon_1,\epsilon_2)
\bar{w}(x',\epsilon_3,\epsilon_4)
\rangle=2(2\pi)^2\delta(\epsilon_1-\epsilon_4)
\delta(\epsilon_2-\epsilon_3) \D(x,x',\epsilon_1-\epsilon_2)
\nonumber \, ,\\&& \langle
w(x,\epsilon_1,\epsilon_2)w(x',\epsilon_3,\epsilon_4)\rangle=-g(2\pi)^3\delta(\epsilon_1-\epsilon_4)
\delta(\epsilon_2-\epsilon_3) \int dx_1
\D_{\epsilon_1-\epsilon_2,x,x_1}\nabla F_{\epsilon_2,x_1} \nabla
F_{\epsilon_1,x_1} \D_{\epsilon_2-\epsilon_1,x_1,x'} \,\,
,\nonumber \\&& \langle \bar{w}(x,\epsilon_1,\epsilon_2)
\bar{w}(x',\epsilon_3,\epsilon_4) \rangle=0 \,\, .
\end{eqnarray}
\begin{multicols}{2}
It is the the third cumulant (the reduced correlation function) of
the current, $S_3$,  that is of interest to us
\begin{eqnarray}&&
\label{z38} {\cal{S}}_3=\S_3-3I\S_2+2I^3\, \, .
\end{eqnarray}

To evaluate  $\S_3$ one follows steps similar to those that led to
the derivation of $\S_2$, see Ref. \cite{GG}. If all relevant
energy scales  in the problem are smaller than the transversal
Thouless energy ($E_{Th}=D/L_T^2$, where $L_T$ is a width of a
wire), the wire is effectively quasi-one dimensional. In that case
only the lowest transversal mode of the diffusive propagator can
be taken into account,
 which yields
\begin{eqnarray}&&
\D(x_1,x_2)=\frac{1}{2\pi
g}\bigg[|x_1-x_2|-x_1-x_2+\frac{2x_1x_2}{L}\bigg]\, \, .
\end{eqnarray}
Here $g=\hbar\nu D$. The electron distribution function in this
system is equal to
\begin{eqnarray}&&
\label{distribution_function}
F(x,\epsilon)=\frac{x}{L}F_{eq}\left(\epsilon-\frac{eV}{2}\right)+
\left(1-\frac{x}{L}\right)F_{eq}\left(\epsilon+\frac{eV}{2}\right)\,\,.
\end{eqnarray}
The quantities $F$ and $\D$ determine the correlation functions,
eq.~(\ref{corr1}). We can now begin to evaluate $\S_3$, (c.f.
eq.~(\ref{z28})), performing a perturbative expansion in the
fluctuations around the saddle point solution, eq.~(\ref{e25}).
After some algebra we find that in the zero frequency limit the
third order {\it cumulant}  is given by



\begin{eqnarray}&&
\label{a10}
{\cal{S}}_3(\omega_1=0,\omega_2=0)=\frac{3e^3{\cal{A}}\pi
g^2}{\hbar L^3} \int_0^L d x_1 dx_2 \nonumber \\&&
\int_{-\infty}^{\infty}d\epsilon F(\epsilon,x_1) \D[0,x_1,x_2]
\nabla\left(F^2(\epsilon,x_2)\right) \,\, .
\end{eqnarray}
Integrating over energies and coordinates we obtain
\begin{eqnarray}&&
\label{e27} {\cal{S}}_3(\omega_1=0,\omega_2=0)=e^2 I y(p)\,\, ,
\nonumber \\&&
y(p)=\frac{6(-1+e^{4p})+(1-26e^{2p}+e^{4p})p}{15p(-1+e^{2p})^2}
\,\, ,
\end{eqnarray}
where $p=eV/2T$. The function $y$ is depicted in Fig.1, where it
is plotted on a logarithmic scale.
\begin{figure}[h]
\includegraphics[width=0.45\textwidth]{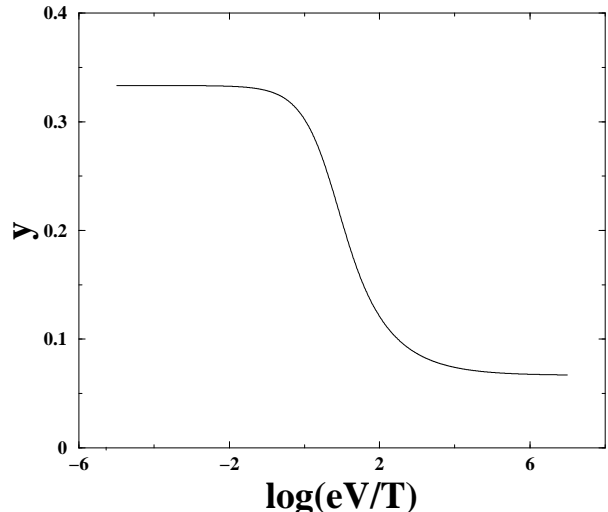}
\label{fig1} \caption{The scaling function $y$ plotted on a
logarithmic scale, cf. eq.~(\ref{e27})}
\end{figure}

Let us now discuss the main features of the function $S_3$. In
agreement with symmetry requirements ${\cal{S}}_3$ is an odd
function of the voltage (the even correlator ${\cal{S}}_2$ is
proportional to the absolute value of voltage), and vanishes at
equilibrium. The zero temperature result (high voltage limit) has
already been obtained by means of the scattering states approach
for single-channel systems \cite{Levitov_Lesovik}, and later
generalized by means of Random Matrix Theory (RMT) to
multi-channel systems (chaotic and diffusive) \cite{Levitov_Lee}.
In our derivation we do not assume the applicability of RMT. Our
result covers the whole temperature range. We see that at low
temperatures the third order cumulant  is linear in the voltage
\begin{eqnarray}&&
\label{a11} {\cal{S}}_3=\frac{e^2}{15} I\,\, .
\end{eqnarray}
At high temperatures the electrons in the reservoirs are not
anymore in the ground state, so the correlations are partially
washed out by thermal fluctuations. One may then expand $y(p)$,
eq.~(\ref{e27}), in a  series of $eV/2T$. The leading term in this
high temperature expansion  is linear in the voltage
\begin{eqnarray}&&
\label{a12} {\cal{S}}_3=\frac{e^2}{3} I\,\, .
\end{eqnarray}
Note that although thermal fluctuations enhance the noise
(compared with the zero temperature limit), eqs.(\ref{a11}) and
(\ref{a12}) differ by only a numerical coefficient. For the
symmetry reasons alluded to in the introduction (the vanishing of
${\cal{S}}_3,\,\, {\cal{S}}_5$, etc. at equilibrium),
 it is tempting to
conjecture  that  all odd order current cumulants interpolate
between two constant values as the ratio  $eV/kT$ is varied. The
experimental study of these cumulants  provides one with a direct
probe of non-equilibrium behavior, not masked by equilibrium
thermal fluctuations.

\section{Inelastic Case}
\label{sec:inelastic} In our analysis so far we have completely
ignored  inelastic collisions among the electrons. This procedure
is well justified  provided that the inelastic length greatly
exceeds the system's  size. However, if this is not the case,
different analysis is called for.  To understand why inelastic
collisions do matter for  current fluctuations, we would like to
recall the analysis of ${\cal{S}}_2$ for a similar problem.
 The latter function is fully determined by the effective electron
temperature. Collisions among  electrons, which are subject to an
external bias,  increase the temperature of those electrons. This,
in turn, leads to the  enhancement of ${\cal{S}}_2$, cf. Refs.
\cite{Nagaev95,Kozub}. In the limit of short inelastic length
\begin{equation}
\label{inelastic} l_{\rm in} \ll L  \,\, ,
\end{equation}
the zero frequency and zero temperature noise is

\begin{equation}
{\cal{S}}_2(0)=\frac{\sqrt{3}}{4}eI \,\, .
\end{equation}
By comparing with the $\omega, T=0$ limit of eq.~(\ref{cc1}) we
see that  inelastic collisions enhance the current fluctuations.
In the present section we consider the effect of inelastic
electron collisions on
 ${\cal{S}}_3$.

In the present analysis we  assume that the electron-phonon
collision length is large, $l_{\rm e-ph} \gg L$, hence
electron-phonon scattering may be neglected. The Hamiltonian we
are concerned with is
\begin{eqnarray}&&
H=H_0+H_{\rm int}.
\end{eqnarray}
The Coulomb interaction among the electrons is described by
\begin{eqnarray}&&
H_{\rm int}=\frac{1}{2} \int d{\bf r} d{\bf r'} \bar\Psi({\bf r})
\bar\Psi({\bf r}') V_0({\bf r}-{\bf r}') \Psi({\bf r})\Psi({\bf
r}')\,\, ,
\end{eqnarray}
where
\begin{eqnarray}&&
V_0({\bf r}-{\bf r}')=\frac{e^2}{|{\bf r}-{\bf r}'|} \,\,.
\end{eqnarray}
 We need to deal with the  effect of electron-electron
interactions  in the presence of disorder and away from
equilibrium. Following Ref. \cite{KA} one may introduce an
auxiliary bosonic field
\begin{eqnarray}&&
\Phi=\left(
\begin{array}{l}
\phi_1 \\
\phi_2
\end{array} \right),
\end{eqnarray}
which  decouples the interaction in the particle-hole channel. Now
the partition function (eq.  (\ref{gen1})) is a functional
integral over both  the  bosonic fields $Q$ and $\phi$\, ,
\begin{eqnarray}&&
\label{gen2} \langle Z \rangle =\int_{Q^2=1} {\cal D}Q {\cal
D}\phi\exp(iS_{total})\, \, .
\end{eqnarray}
The action is
\begin{mathletters}
\label{act1,act2,act3}
\begin{eqnarray}&&
iS_{total}=iS[\Phi]+iS[\Phi,Q] \, , \label{act1} \\&&
iS[\Phi]=i\Tr\{\Phi ^T V^{-1}_0\gamma^2\Phi\}  \, , \label{act2}
\\&& iS[\Phi,Q]=-\frac{\pi\nu}{4\tau}\Tr\{Q^2\}\!+ \Tr\ln
\Big[\hat G_0^{-1}\!+\frac{iQ}{2\tau}+\phi_{\alpha}\gamma^{\alpha}
\Big].  \label{act3}
\end{eqnarray}
\end{mathletters}
It is convenient to perform a ``gauge transformation'' \cite{KA}
to a new field $\tilde{Q}$
\begin{equation}
\label{g1} Q_{t,t'}(x) = \exp\left(i
k_{\alpha}(x,t)\gamma^{\alpha}\right)
 \tilde Q_{t,t'}(x)
\exp\left(-i k_{\alpha}(x,t')\gamma^{\alpha}\right)\, .
                                                              \label{p9a}
\end{equation}
Introducing the long derivative
\begin{equation}
\label{long_derivative} {\bf \partial}_x\tilde Q \equiv \nabla
\tilde Q +i [\nabla k_{\alpha}\gamma^{\alpha}, \tilde Q]\, ,
                                                            \label{q9}
\end{equation}

one may write the gradient expansion of   eq.~(\ref{act3}) as
\begin{eqnarray}
&&iS[\tilde Q,\Phi]= i\nu \Tr \{(\Phi-i\omega K)^T \gamma_2
(\Phi+i\omega K) \} -
                                                             \label{q8}\\
&&\frac{\pi\nu}{4} \left[ D  \Tr \{ {\bf \partial}_x\tilde Q\}^2 +
4i \Tr \{(\epsilon +(\phi_\alpha+i\omega
k_{\alpha})\gamma^{\alpha}) \tilde Q \} \right] \, \, .
                                                                \nonumber
\end{eqnarray}
At this point the  vector $K^T=(k_1,  k_2)$ that determines the
transformation (\ref{p9a}) is arbitrary. The saddle point equation
for $Q$  of the action (\ref{q8})
 is given by the  following equation
\begin{equation}
\label{sp2} D \partial_x ( \tilde{Q} \partial_x\tilde{Q} ) + i
[(\epsilon +(\phi_\alpha+i\omega k_{\alpha})\gamma^{\alpha}),
\tilde{Q} ]  = 0 \, .
                                                            \label{k6}
\end{equation}
Let us now choose the parameterization
\begin{equation}
\label{par2} \tilde{Q}=\tilde{\Lambda}\exp(\tilde{W}),
\end{equation}
where $\tilde{W}$ represents fluctuation around the saddle-point
\begin{eqnarray}&&
\label{ansatz} \tilde{\Lambda}(x,\epsilon)=\left(\matrix{1 &
2\tilde{F}[\phi](x,\epsilon) \cr 0 & -1\cr}\right)\,\, .
\end{eqnarray}
Eq. (\ref{ansatz}) implies that the solution of the  saddle point,
equation (\ref{sp2}), determines $\tilde{F}$ as a functional of
$\phi$. We do not know, though,  how to solve it. Instead we
average  over $\phi$ on the level of eq.(\ref{sp2}). The solution
of this averaged equation,  denoted by $\bar{F}$, is determined
by:
\begin{equation}
\label{c3} D\nabla^2 \bar{F}(\epsilon)=I^{ee}\{F\}\, ,
\end{equation}
where the r.h.s. is given by
\begin{eqnarray}&&
\label{e29} I^{ee}\{\bar{F}\}=D \int \frac{d\omega}{\pi}\big[
\langle \nabla k^1(\omega)\nabla\!k^1(-\omega)\rangle
\big(\bar{F}(\epsilon)-\bar{F}(\epsilon-\omega)\big)+ \nonumber
\\&&
\!(\langle\nabla\!k^1(\omega)\nabla\!k^2(-\omega)\rangle\!-\!\langle\nabla\!k^2(\omega)\nabla\!k^1(-\omega)\!\rangle)
\big(\bar{F}(\epsilon)\bar{F}(\epsilon\!-\omega)\!-\!1\big)\big].
\end{eqnarray}
We stress that our averaging procedure is not related to the
genuine interacting saddle point. We will be expanding about
another point ($\bar{\Lambda}$) at which the fields $\tilde{W}$
and $\phi,(\nabla k)$ are nearly decoupled; $\tilde{W}$ is
parameterized as in eq.(\ref{c20}). Taking variation of the action
with respect to $w$, $\bar{w}$, we obtain the following gauge,
determining $k[\phi]$:
\begin{eqnarray}&&
\label{gauge} D\nabla^2 k_2-\phi_2-i\omega k_2=0 \nonumber \\&&
D\nabla^2k_1+\phi_1+i\omega k_1=2\B[\omega,x]\nabla^2 k_2 \, ,
\end{eqnarray}
where
\begin{eqnarray}&&
\B[\omega,x]=\frac{1}{2\omega}\int
d\epsilon[1-\bar{F}(\epsilon,x)\bar{F}(\epsilon-\omega,x)]\, .
\end{eqnarray}
Though we have failed to find the true saddle point  the linear
part of the action expanded around ($\bar{\Lambda}$) is zero. It
is remarkable to notice that under conditions (\ref{gauge})
eq.(\ref{c3}) becomes a quantum kinetic equation
\cite{quantum_kinetic_equation} with the  collision integral being
$I^{ee}\{F\}$. The above procedure consists of an expansion of the
action, eq.(\ref{q8}), around its saddle point. The expansion
presumes slow fluctuations in space and time (though the results
obtained, expressing ${\cal{S}}_3$ in terms of the  distribution
function and the effective propagator, eq.(\ref{a10_inelastic}),
may be valid under less restrictive conditions). In  Fourier space
(with respect to the spatial coordinate and the sum--rather than
the difference-- of the time coordinates) this implies $ql \ll 1$
and $\omega\tau \ll 1$.  Hence this procedure  implies that the
collision integral,  $I^{ee}$, appearing in the quantum kinetic
equation, eq.(\ref{c3}), accounts only  for electron-electron
collisions with small momentum transfer. We therefore need to
ascertain that these are indeed the small momentum transfer
processes which make the dominant contribution to the
electron-electron collision rate. To estimate  the relative
importance of the small and large momentum  processes we compare
their respective contributions to the energy relaxation time,
$\tau_{ee}$.
 Electron-electron scattering with small momentum transfer  leads to
result \cite{quantum_kinetic_equation,AA}:
\begin{equation}
1/\tau_{ee} = \frac{\epsilon^{d/2}}{\hbar\nu_d D^{d/2}} \,\,\, \,
ql \ll 1\, \, ,
\end{equation}
while the contribution coming from large momentum transfer is
\begin{equation}
1/\tau_{ee}=\frac{\pi^2}{64}\frac{\epsilon^2}{\mu}\frac{\kappa}{p_F}
\,\,\,\,  ql \gg 1\, \, .
\end{equation}
As is now clearly seen, for low enough energies
\begin{equation}
\epsilon  \ll \left(\frac{64\mu
p_F}{\pi^2\kappa\hbar\nu_dD^{d/2}}\right)^{\frac{2}{4-d}}\, \, ,
\end{equation}
the contribution to the inelastic mean free time coming from
scattering with small momentum transfer  is dominant; ignoring it
would result in seriously overestimating $\tau^{ee}$ time (and
correspondingly  the mean free path $l_{\rm ee}$).

Coming back to our calculations we note that the correlation
function of current fluctuations is a gauged invariant quantity
(does not depend on the position of the Fermi level). This means
that momenta $q \le \sqrt{\omega/D}$ do not contribute to   such a
quantity \cite{Fin}. In this case the Coulomb propagator is
universal, i.e. does not depend on the electron charge. The fact
that we address gauge invariant quantities  allows us to represent
the generating functional $Z$ in terms of the fields $Q$ and
$\nabla k$ (rather than $Q$ and $\phi$), as in Ref \cite{KA}.
\end{multicols}
\begin{equation}
\label{c22} \langle Z \rangle \! = \! \int\!\! {\cal D}\nabla K\,
\exp\left( -i\nu D \Tr\{ \nabla K^T {\cal D}^{-1} \nabla
K\}\right) \int {\cal D} \tilde Q\, \exp\left(
\sum\limits_{l=0}^{2} i S_{l}[\tilde Q,\nabla K]\right) \, .
                                                              \label{u21}
\end{equation}
Here we define
\begin{equation}
{\cal D}^{-1} = \left(
\begin{array}{cc}
0 & -D\nabla^2_x  + i\omega \delta_{x,x'}  \\
-D\nabla^2_x  -  i\omega \delta_{x,x'} & -2i\omega
\delta_{x,x'}B_\omega(x)
\end{array}
\right)  \, \, ,
                                                              \label{k15}
\end{equation}
where  the  expansion $S=S^0+S^1+S^2$, is in powers of  $\nabla
K$; the $l-th$ power  ($ l=0,1,2$)  is
 given by
\begin{mathletters}
\label{q11,q12,q13}
\begin{eqnarray}
&&iS^0[\tilde Q]= -\frac{\pi\nu}{4} \left[ D  \Tr \{ \nabla \tilde
Q\}^2 +  4i\,  \Tr \{\epsilon \tilde  Q \} \right]\, ,
                                                              \label{q11} \\
&&iS^1[\tilde Q,\nabla K]= -i\pi\nu \left[ D  \Tr \{ \nabla
k_\alpha\gamma^{\alpha} \tilde Q\nabla \tilde Q \} + \Tr
\{(\phi_\alpha+i\omega  k_\alpha)\gamma^{\alpha} \tilde Q\}
\right]\, ,
                                                              \label{q12} \\
&&iS^2[\tilde Q,\nabla K]= \frac{\pi\nu D}{2} \left[ \Tr \{ \nabla
k_\alpha\gamma^{\alpha} \tilde Q
       \nabla k_\beta \gamma^{\beta } \tilde  Q \} -
\Tr \{ \nabla k_\alpha\gamma^{\alpha} \tilde{\Lambda}
       \nabla k_\beta \gamma^{\beta } \tilde{\Lambda} \}
\right]\, .
                                                              \label{q13}
\end{eqnarray}
\end{mathletters}
From eq.(\ref{c22}) we obtain the  gauge field correlation
function
\begin{equation}
\label{c2} \langle \nabla k_\alpha(x,\omega) \nabla
k_\beta(x',-\omega) \rangle =
\frac{i}{D}Y_{\alpha,\beta}(\omega,x,x'),
\end{equation}
where
\begin{equation}
\label{c4} Y(\omega,x,x')=  \left[ \begin{array}{cc}
- 2i\pi \nu \omega \int d x_1 \D[-\omega,x,x_1]\B[\omega,x_1]\D[\omega,x_1,x']               &\D[-\omega,x,x'] \\
\D[\omega,x,x']  &  0
\end{array} \right] ,
\end{equation}
 Using eqs.(\ref{c2},\ref{c4}) we rewrite eq.(\ref{e29}) for the
quasi-one-dimensional wire as:
\begin{eqnarray}&&
\label{e36} D\nabla^2 \bar{F}(\epsilon)=I^{ee}\{\epsilon,x\}\, ,
{\rm where }  \\&& \label{c7}
I^{ee}(\epsilon,x)=\frac{i\pi}{2}\int d\omega \big[
-2i\omega\pi\nu
D[x,x_1,-\omega]B[\omega,x_1]\D[x_1,x,\omega]\big(\bar{F}(\epsilon)-\bar{F}(\epsilon-\omega)\big)+\nonumber
\\&&
(\D[x,x,\omega]-\D[x,x,-\omega])(1-\bar{F}(\epsilon)\bar{F}(\epsilon+\omega))
\big].
\end{eqnarray}
\begin{multicols}{2}
The total number of particles and the total energy of the
electrons are both preserved during electron-electron and  elastic
electron-impurity scattering.  The collision integral,
eq.(\ref{c7}),  satisfies then
\begin{eqnarray}&&
\label{c8} \int_{-\infty}^{\infty}I^{ee}(\epsilon,x)d\epsilon=0,
\end{eqnarray}
\begin{eqnarray}&&
\label{c9} \int_{-\infty}^{\infty}\epsilon
I^{ee}(\epsilon,x)d\epsilon=0.
\end{eqnarray}
We now consider the limit   $l_{\rm ee} \ll L$. The solution of
eq.(\ref{c3}) assumes then the  form of a quasi-equilibrium
single-particle distribution function
\begin{eqnarray}&&
\label{c12} \bar{F}(\epsilon,x)=\tanh\left({\frac{\epsilon
-e\phi(x)+\bar{\mu}}{2T(x)}}\right).
\end{eqnarray}
Here $\epsilon$ is the total energy of the electron, and $e\phi$
is the local electro-chemical potential, i.e. the sum of the local
chemical potential $\mu$ and the electrostatic potential $u$
($e\phi(x)=\mu(x) + eu(x)$); $\bar{\mu}$ is the  value of the
electro-chemical potential before the bias has been applied and
$T(x)$ is the effective local  temperature of the electron gas.

Since the size of a constriction is much longer than Debye
screening length $\lambda^{-1}_D=\sqrt{4\pi e^2\nu}$ we may use a
quasi-neutrality approximation, in this case to assume that a
value of chemical potential $\mu$ is constant in the constriction.
It still remains to find the effective electronic temperature,
$T(x)$,  and the electrostatic potential $u(x)$.

In order to find the electrostatic  potential that enters
eq.(\ref{c12}) we employ eq.~(\ref{c8}). To facilitate our
calculations we further  assume that conductance band is symmetric
about the Fermi energy and that the spectral density of
single-electron energy levels is constant.
Integration over the energy, eq.(\ref{c8}) yields\\
\begin{equation}
\label{e30}
\partial_x^2\int \bar{F}(\epsilon)d\epsilon =0
\end{equation}
\begin{equation}
\label{electrostatic} {\partial_x}^2 u=0.
\end{equation}
Solving  eq.(\ref{electrostatic}) with the requirement that the
electro-chemical potential at the edges of the sample differs by
$V$, we find  that the electro-chemical  potential along the
constriction $(0 \le x \le L)$ is given by
\begin{equation}
\label{c6} e\phi(x)=eV\left(\frac{x}{L}-\frac{1}{2}\right)
+\bar{\mu}.
\end{equation}

We use the energy conservation property of the collision integral,
eq.~(\ref{c9}). Multiplying  eq.~(\ref{e36}) by energy and
integrating over it
\begin{equation}
\label{c5}
\partial_x^2\int \epsilon \bar{F}(\epsilon)d\epsilon =0\, \, ,
\end{equation}
we obtain an  equation for the electron temperature:
\begin{equation}
\label{eq_for_temperature}
\partial_x^2\bigg(\frac{\pi^2}{6}(kT(x))^2+\frac{1}{2}(e u(x))^2\bigg)=0.
\end{equation}
 The boundary condition of eq.(\ref{eq_for_temperature}) is
determined by the temperature of the electrons in the reservoirs.
Combining eqs.(\ref{c5})~and~(\ref{c6}) we find the electron
temperature in two opposite limits:
\begin{eqnarray}&&
\label{temperature} T(x)=\left\{
\begin{array}{l}
\frac{\sqrt{3}eV}{\pi L}\sqrt{x(L-x)}\,\,\,\,\,\,  eV \gg T\,\, ,
\\
 \,\,\,\,\, T \hspace{2.3cm}eV \ll T\,\,.
\end{array}
\right.
\end{eqnarray}
Eqs.~ ((\ref{c12})\,, (\ref{c6})\,  and (\ref{temperature})
determine the function $\bar{F}$ uniquely. We now replace the
right-corner element of the matrix $\tilde{\Lambda}$ (i.e.
$\tilde{F}[\phi]$, cf. eq.(\ref{ansatz})) by its average value
$\bar{F}$.

To calculate ${\cal{S}}_3$ under conditions of strong
electron-electron scattering (eq.~(\ref{inelastic})) one needs to
replace the operators
 $\hat{\I}^D$ and $\hat{\M}$  in eq.~ (\ref{z28})
by their gauged values
\begin{eqnarray}&&
\label{S3_inel} \S_3(t_1-t_2,t_2-t_3)=\frac{e^3(\pi\hbar\nu
D)^2}{8}\nonumber \\&& \bigg\langle\frac{1}{2}
\hat{\tilde{\M}}(x_1,t_1)\hat{\tilde{\I}}^D
(x_2,x_3,t_2,t_3)+\nonumber \\&& (x_1,t_1\leftrightarrow\!x_3,t_3)
+ (x_1,t_1\leftrightarrow\!x_2,t_2)+ \nonumber \\&&
\frac{\pi\hbar\nu
D}{8}\hat{\tilde{\M}}(x_1,t_1)\hat{\tilde{\M}}(x_2,t_2)\hat{\tilde{\M}}(x_3,t_3)\bigg\rangle_{\nabla
k, \tilde{Q}} \,\,  ,
\end{eqnarray}
where the averaging is taken over the entire action $S$ and the
Gaussian weight function for $\nabla K$, as in eq.~(\ref{gen2}).
Here we define (cf. eqs.(\ref{e33}), (\ref{M}) with eqs.
(\ref{e332}),(\ref{c21}))
\begin{eqnarray}&&
\label{e332}
\hat{\tilde{\I}}^D(x,x',t,t')=\Tr\!\bigg\{\!\tilde{Q}_{x,t,t'}\gamma_2\tilde{Q}_{x',t',t}\gamma_2-\delta_{t,t'}\gamma_1\bigg\}\delta_{x,x'}
\,\, ,
\end{eqnarray}
\begin{eqnarray}&&
\label{c21}
\hat{\tilde{\M}}(x,t)=\Tr\!\bigg\{\!\int\!dt_1\bigg([\tilde{Q}_{x,t,t_1};\partial_x
]~ \tilde{Q}_{x,t_1,t} \bigg)\gamma_2\bigg\}\,\, ,
\end{eqnarray}
where the ``long derivative'', $\partial_x$, is presented in
eq.(\ref{long_derivative}). In order to actually perform the
functional integration over the matrix field $\tilde{Q}$ we use
the parameterization of eq.~(\ref{par2}). The operators
$\hat{\tilde{\I}}^D,\hat{\tilde{\M}}$ may be expanded over
$w,\bar{w}$ and $\nabla k$ :
\begin{eqnarray}&&
\hat{\tilde{\M}}=\tilde{\M}^0_0+\tilde{\M}^0_1+\dots+\tilde{\M}^1_0+\tilde{\M}^1_1+
\dots \\ && \hat{\tilde{\I}}^D=\tilde{\I}_0+\tilde{\I}_1 + \dots
\end{eqnarray}
Here the upper index refers to the power of the  $\nabla k$
fields; the lower refers to the power  of $w, \bar{w}$ fields in
the expansion. We need  to find the Gaussian fluctuations around
the saddle point of the action (\ref{q11,q12,q13}). Though we did
not find  the exact saddle point, the expansion of  $Q$ around
$\bar{\Lambda}$ works satisfactorily. The coupling between the
fields $\nabla k$ and $W$ which appears already in the Gaussian
(quadratic) part is small, since it is proportional to the
gradient of the distribution function:
\begin{eqnarray}&&
\label{action_out_of_equlibrium} \!iS^1_1\!=\!-2i\pi\!g
\Tr\bigg\{\! \bar{w}_{x,\epsilon,\epsilon'} [\nabla k_{1
x,\epsilon'-\epsilon} \nabla\bar{F}_{x,\epsilon}-
\nabla\bar{F}_{x,\epsilon'}\nabla\!k_{1 x,\epsilon'-\epsilon}\!+
\nonumber \\&& \nabla\bar{F}_{x,\epsilon'}\nabla\!k_{2
x,\epsilon'\epsilon}\bar{F}_{x,\epsilon}+
\bar{F}_{x,\epsilon'}\nabla\!k_{2
x,\epsilon'-\epsilon}\nabla\bar{F}_{x,\epsilon}] \bigg\}.
\end{eqnarray}
Considered as a small perturbation, $iS_1^1$ does not affect the
results.

The more dramatic effect on the correlation function arises from
the non-Gaussian  part of the action, eqs. (\ref{q12},\ref{q13})
(by this we mean non-Gaussian terms in either $w, \bar{w}$ or
$\nabla K$). After integrating over the interaction an additional
contribution to the Gaussian part (proportional to $w\bar{w}$) of
the action arises. To find the effective action $iS^{\rm eff}[W]$
we  average over the interaction along the following lines:
\begin{eqnarray}&&
\label{mean_field}
\bigg\langle\exp\bigg\{iS^1+iS^2\bigg\}\bigg\rangle_{\!\!\!\nabla\!k}
= \nonumber \\&& \bigg\langle\exp\bigg\{\bigg\langle
iS^1+iS^2\rangle + iS^1+iS^2-\bigg\langle iS^1+iS^2\bigg\rangle
\bigg\}\bigg\rangle_{\!\!\!\nabla\!k} \simeq\nonumber
\\&&\!\exp\bigg\{\!\bigg\langle\!iS^1+iS^2\bigg\rangle_{\!\!\!\nabla\!k}
\bigg\}\bullet \nonumber \\&&
\bigg[1\!+\!iS^1+iS^2\!-\!\bigg\langle
\!iS^1\!+\!iS^2\bigg\rangle_{\!\!\!\nabla\!k}\!\!\!+\!
\frac{1}{2}\!\bigg(\!iS^1\!+\!iS^2\!-\!\bigg\langle\!iS^1\!+\!iS^2\!\bigg\rangle_{\!\!\!\nabla\!k}\!\!\bigg)^2\!\bigg]\nonumber
\\&& \simeq
\!\!\exp\bigg\{\bigg\langle\!iS^1+iS^2\!\bigg\rangle_{\!\!\!\nabla\!k}\!\!\!\!+\!\!\!\frac{1}{2}\!\bigg\langle\!\bigg(iS^1+iS^2\bigg)^2\bigg\rangle_{\!\!\!\nabla\!k}\!\!\!\!\!-\!\!\frac{1}{2}\!\bigg\langle\!iS^1\!+\!iS^2\bigg\rangle_{\!\!\!\nabla\!k}^2\!\!\bigg\}
.
\end{eqnarray}
There is no linear term in $\nabla k$ and the Gaussian part has
been separated  out the following identity hold:
\begin{eqnarray}&&
\label{identity1} iS_0^1=iS_0^2=0\, ,
\end{eqnarray}
(where, again, $S_0^1$ refers to the component of the action,
eq.(\ref{q11,q12,q13}), that has  zero power of the $w,\bar{w}$
fields and one power of the $\nabla k$ field). In addition, due to
the choice of the gauge, eq.(\ref{gauge}), and the condition
($l_{\rm ee} \ll L$), the averaging over $\nabla k$ does not
generate  terms linear in $w,\bar{w}$ in the effective action:
\begin{eqnarray}&&
\label{identity2} \langle iS^2_1\rangle_{\nabla
k}=-2i\pi\nu\int\frac{d\epsilon}{2\pi}\bar{w}_{\epsilon,\epsilon}
I_{\rm ee}[F]=0\, .
\end{eqnarray}
Combining eqs. (\ref{mean_field},\ref{identity1} and
\ref{identity2}) we find that the effective action acquires an
additional contribution:
\begin{eqnarray}&&\hspace{0.4cm}
\bigg\langle \exp\left(iS^1+iS^2\right)\bigg\rangle_{\nabla
k}\simeq\exp\left(\langle iS^2_2\rangle+\frac{1}{2}\langle iS^2_1
iS^2_1\rangle\right)
\end{eqnarray}
\end{multicols}
The general form of the effective action is rather complicated,
however for the low frequency noise only diagonal part of the
action matters:
\begin{eqnarray}&&
\label{e55}iS^{\rm eff}_2[w,\bar{w}]\!=
\frac{\pi\nu}{2}\Tr\bigg\{\bar{w}_{x,\epsilon,\epsilon}
\!\bigg[\!-D\nabla^2\!+\!\hat{{\cal{I}}}^{ee}\bigg]w_{x,\epsilon,\epsilon}-
\bar{w}_{x,\epsilon,\epsilon}D\nabla\bar{F}_{x,\epsilon}\nabla\bar{F}_{x,\epsilon}\bar{w}_{x,\epsilon,\epsilon}
\bigg\}\, \, .
\end{eqnarray}
Here the operator
\begin{eqnarray}&&
\label{linearized_col_int}
\hat{{\cal{I}}}^{ee}w_{x,\epsilon,\epsilon}\equiv \int d
\omega[Y_{11}(\omega)[w_{\epsilon,\epsilon}-w_{\epsilon-\omega,\epsilon-\omega}]+\nonumber
\\&& \left(Y_{12}(\omega)-Y_{21}(\omega)\right) [F_\epsilon
w_{\epsilon-\omega,\epsilon-\omega}+F_{\epsilon-\omega}w_{\epsilon,\epsilon}]+\nonumber
\\&& \int d\bar{\epsilon}\frac{1}{2\omega}
\left(F_{\epsilon}-F_{\epsilon-\omega}\right)
\left(Y_{12}(\omega)-Y_{21}(\omega)\right)
\left(F_{\bar{\epsilon}+\omega}+F_{\bar{\epsilon}-\omega}\right)
w_{\bar{\epsilon},\bar{\epsilon}}
\end{eqnarray}
is a {\it linearized} collision integral, i.e. a variation of the
collision integral (\ref{c7}) with respect to the distribution
function. Inspecting eq.(\ref{S3_inel}) we find (details are
outlined in Appendix \ref{appendix1})
\begin{eqnarray}&&
\label{a10_inelastic}
{\cal{S}}_3(\omega_1=0,\omega_2=0)=\frac{3e^3{\cal{A}} \pi
g^2}{\hbar L^3} \int_0^L d x_1 dx_2 \nonumber \\&&
\int_{-\infty}^{\infty}d\epsilon_1d\epsilon_2
\bar{F}(\epsilon_1,x_1) {\cal{D}}[x_1,\epsilon_1;x_2,\epsilon_2]
\nabla\left(\bar{F}^2(\epsilon_2,x_2)\right) \, \, .
\end{eqnarray}
The diffusion propagator in the presence of strong inelastic
electron-electron scattering is given by \cite{GGM}
\begin{eqnarray}&&
\label{inelasticd}
{\cal{D}}[x_1,\epsilon_1;x_2,\epsilon_2]=\left(\frac{\partial}{d\epsilon}
f_0\left(\frac{\epsilon_1-\mu(x_1)}{T(x_1)}\right)\right)\D[x_1,x_2]\left(-1-
\frac{3}{\pi^2} \frac{\epsilon_1-\mu(x_1)}{T(x_1)}
\frac{\epsilon_2-\mu(x_1)}{T(x_1)} \right)\, ,
\end{eqnarray}
where
\begin{eqnarray}&&
 f_0(x)=\frac{1}{1+\exp(x)}.
\nonumber
\end{eqnarray}
 Evaluating ${\cal{S}}_3$ explicitly we find that the third
order current cumulant is
\begin{eqnarray}&&
{\cal{S}}_3(\omega_1=0,\omega_2=0)=\frac{36e^3Ag^2
eV}{L^4\pi}\int_0^L dx_1dx_2
D[x_1,x_2]\bigg[\frac{T(x_1)}{T(x_2)}+(x_1-x_2)\frac{1}{T(x_2)}\frac{\partial}{\partial
x_1} T(x_1)\bigg].
\end{eqnarray}
At high temperatures (cf. eq.\ref{temperature}) one obtains
\begin{eqnarray}&&
{\cal{S}}_3(\omega_1=0,\omega_2=0)=\frac{3}{\pi^2}e^2I \, ,
\end{eqnarray}
while at low temperatures
\begin{eqnarray}&&
{\cal{S}}_3(\omega_1=0,\omega_2=0)=\left(\frac{8}{\pi^2}-\frac{9}{16}\right)
e^2I .
\end{eqnarray}

Our calculation was performed  for a simple rectangular
constriction. However, our results hold for any shape of the
constriction, provided it is quasi-one dimensional (we have
considered a single transversal mode only).

Finally we discuss the role of an applied magnetic field. Consider
a two-terminal geometry and a non-interacting electron gas.  For
such a system any  current correlation function  can be expressed
through the channel transmission probabilities $\{T_i\}$.
Onsager-type relations dictate that the transmission probabilities
are invariant under the reversal of the sign of the magnetic flux
(``phase locking''), i.e., $T_i(\phi)=T_i(-\phi)$. This   implies
${\cal{S}}_3(\phi)={\cal{S}}_3(-\phi)$. Let us take the zero bias
voltage limit. Because ${\cal{S}}_3$ is an even function of the
magnetic flux   it cannot change its  sign under time-reversal
operation.  On the other hand, by its very definition, it must
reverse its sign under the time reversal operation. It therefore
must vanish. We conclude that the  magnetic field alone, while
breaking the time reversal symmetry,  cannot lead to finite-value
odd current correlators.

\section{Why The Standard Kinetic Equation Approach Fails}
\label{sec:kinetic} The phase coherence is not significant for the
current fluctuations in the weakly disordered metals (the
statement remain true for any system for which electron dynamics
is classical). This suggest that the kinetic theory of
fluctuations may be suitable description of the problem. Here we
state, that even though the dephasing length may be short ($l_{rm
ph} \ll L$) the is no purely classical route to find high order
current correlation functions.

One possible way to represent the kinetic theory of fluctuations
is through Boltzmann-Langevin equation \cite{Shulman}. The
deviation $\delta f$ of the exact distribution function from its
coarse-grained) value ($\bar{f}$) satisfies a stochastic equation
with the random  additive noise ($\delta J$):
\begin{eqnarray}&&
\label{Boltzmann_Langevin} \left(\frac{\partial}{\partial t}+{\bf
v}\cdot{\bf \nabla}\right)\delta {f}({\bf p},{\bf r},t)= {\rm
Col}\{f\}+ \delta J^{ext}({\bf p},{\bf r},t).
\end{eqnarray}
The l.h.s. of  eq.~(\ref{Boltzmann_Langevin}) is the conventional
drift term of the kinetic equation;  ${\rm Col}$ is a collision
integral. The random flux term $\delta J$  which  is added to the
standard Boltzmann equation,  induces  fluctuations of the
distribution function $f$. The correlation of the random fluxes
can be related in the universal manner to the coarse-grained
particle flux in the phase space (which is specific for each given
problem).  However the very applicability of the kinetic equation
as well as the applicability of kinetic theory of fluctuation are
controlled by the same parameter (in our case $\epsilon_F\tau$).
However, the normalized ratio between high and low order current
cumulants is small by exactly the same parameter.

To see, where the kinetic (Kogan-Shulman) procedure fails we try
to go along the standard way (for the simplicity for
non-interacting electrons). In the diffusive approximation one may
keep only the lowest harmonics:
\begin{eqnarray}&&
\label{diffusive_approximation1} \delta f({\bf p},{\bf r},t
)=\delta f_0\left (p,{\bf r},t \right)+ {\bf n} \cdot \delta{\bf
f_1}\left (p,{\bf r},t\right),
\end{eqnarray}
\begin{eqnarray}&&
\label{diffusive_approximation2} \delta J^{ext}({\bf p},{\bf
r},t)= \delta J_0(p, {\bf r},t)+{\bf n} \cdot \delta {\bf J}_1(p,
{\bf r},t),
\end{eqnarray}
where ${\bf p}\simeq{\bf n}|p_F|$ (here ${\bf n}$ is  unit vector
in the direction of the momentum). Here ${\bf f}_1$ and ${\bf
J}_1$ are vectors in the direction of the average current (the
x-axis). From
eqs.~(\ref{diffusive_approximation1},\ref{diffusive_approximation2})
one finds
\begin{eqnarray}&&
\delta {\bf f}_1=-v_F \tau\nabla \delta f_0(p,{\bf r})-\tau\delta
{\bf J}_1(p, {\bf r},t).
\end{eqnarray}
In the low frequency limit  particle conservation guarantees that
current  fluctuations through any given cross section are the
same. This allows us to compute  the current fluctuations
integrated over the entire  sample length,
\begin{eqnarray}&&
\delta I(t)=\frac{e\nu v_F\tau}{3L}\int d {\bf r}\int d\epsilon
\delta {\bf J_1}_x(\epsilon,{\bf r},t).
\end{eqnarray}
In order to calculated the third order current cumulant
${\cal{S}}_3(0,0)$ it is crucial to know the zero frequency limit
of the third order correlation function of the random flux.
\begin{eqnarray}&&
{\cal{R}}_3({\bf r}_1,{\bf r}_2,{\bf
r}_3,\epsilon_1,\epsilon_2,\epsilon_3)\!\equiv\int
d(t_1-t_2)d(t_1-t_3)\nonumber \\&& \langle \delta{\bf
J_1}_x(\epsilon_1,{\bf r}_1,t_1) \delta{\bf J_1}_x(\epsilon_2,{\bf
r}_2,t_2)\delta{\bf J_1}_x(\epsilon_3,{\bf r}_3,t_3)\rangle .
\end{eqnarray}
If one would go along the standard assumption (of Poissonian, and
$\delta$-correlated in space random flux), one will get a wrong
answer (smaller as $(l/L)^2$ than the correct one). Instead one
can show, that by choosing the correlation function to be long
correlated:
\begin{eqnarray}&&
\label{correlator_random_fluxes}
{\cal{R}}_3({\bf\!r}_1,{\bf\!r}_2,{\bf\!r}_3,\epsilon_1,\epsilon_2,\epsilon_3)\!=
\!\frac{27\pi}{v_F(\tau\nu)^2}\delta(\epsilon_1\!-\!\epsilon_2)\delta(\epsilon_1\!-\!\epsilon_3)
\delta({\bf\!r}_1-{\bf\!r}_2)\nonumber \\&& \delta(y_1-y_3)
\delta(z_1-z_3) (\nabla\!F^2(\epsilon_1,x_1)) \D(x_1,x_3)
F(\epsilon_1,x_3)
\end{eqnarray}
 eq.~(\ref{a10}) (and similarly eq.~\ref{a10_inelastic})) are reproduced.
Here the function $F (\bar{F})$ is related to the singe-particle
distribution function (cf.  eq.~(\ref{e31})).
Eq.(\ref{correlator_random_fluxes}) shows that the correct
high-order correlation functions of the random fluxes are
non-local in space. This is clearly in contradiction with the
assumption made within the Kogan-Shulman formalism.

In other words, one usually employs the
Boltzmann-Langevin-Kogan-Shulman approach  approximating the
actual probability distribution function of the random current
flux  by a Gaussian distribution. Seemingly, this  is justified
because of the large dimensionless conductance $g \gg 1$. In the
scattering states picture, this condition is equivalent to a large
number of transmission channels. We know \cite{Levitov_Lesovik}
that each channel provides its own independent contribution to the
current noise. According to the central limit theorem, a process
that consists of a large number of independent random
contributions is characterized by a (nearly) Gaussian distribution
function, independently of the distribution function of the
individual contributions involved. However, because the value of
the dimensionless conductance is indeed large, yet finite, the
distribution function of the random current flux deviates from
Gaussian. Such deviations are {\it beyond} the validity of the
Boltzmann-Langevin approach. ${\cal{S}}_3$ is the lowest order
current cumulant that probes these deviations.

This is not surprising. Indeed, since the distribution function
$f$ is a macroscopic quantity, (at equilibrium) it must satisfy
the Onsager regression hypothesis \cite{LL}. The kinetic theory of
fluctuations can be viewed as an extension of the this hypothesis
for a non-equilibrium situation \cite{Lax}. Since the Onsager
hypothesis is restricted to  pair correlation functions only, it
is natural to expect that the kinetic theory has the same
limitations.

\section{Discussion}
\label{sec:discussion} We have noted that due to general {\it
symmetry} reasons odd order cumulants of the  current must be odd
functions of the applied voltage and must therefore vanish at
equilibrium. Provided these cumulants are analytic functions of
$V$, they must increase linearly with  voltage bias when the
latter is small. they increase linearly with the latter. Thus,
contrary to even order cumulants, they are not masked by a
background thermal noise. Even more remarkably, at high
temperature, the third order current cumulant  approaches a
constant value. This is in contradistinction with the  second
order current correlator which diverges with the temperature. This
is the reason why quantities such as ${\cal{S}}_3$ are  suitable
for the study of non-equilibrium current noise even at relatively
high temperatures.

 Our results for ${\cal{S}}_3$, pertaining to both  the elastic and the
 inelastic cases (the high and low temperature limits),  are
  summarized in  Table \, \ref{table1}:
\vspace{0.5cm}
\noindent\begin{center}\begin{minipage}{10cm}
\begin{table}
\noindent\begin{center}\begin{minipage}{5cm}
\begin{tabular}{|r|r|l|}
\hline
  &  $T \ll eV$ & $T \gg eV$ \\
\hline
$l_{\rm{in}}\gg L$ \  & \ $1/15 \ e^2I$& \  $1/3 \ e^2I$ \\
\hline
$l_{\rm{in}}\ll L$ \  & \ $0.248 \ e^2 I$& \ $0.304 \ e^2I$ \\
\hline
\end{tabular}\end{minipage}\end{center}
\caption{Limiting values of ${\cal{S}}_3$. }
  \label{table1}
\end{table}\end{minipage}\end{center}
\vspace{0.5cm}

In the elastic limit one can regard the electrons as
non-interacting; the problem then is a natural extension of the
ballistic multichannel setup. Indeed, at low frequencies one may
ignore the  dynamics of the electrons within  the constriction,
which allows us to employ the Landauer scattering states approach
\cite{Landauer}. The latter  consists of describing the
constriction by a quantum transmission matrix, expressing the
currents and their correlation functions through the transmission
coefficients. (It has been noted, though, that coherency is not
essential for deriving the leading order contributions to
shot-noise in disordered systems, cf. Refs.
\cite{deJong_Beenakker,Nagaev2}).

Since odd order current correlation functions vanish at
equilibrium, they  must be proportional to the difference between
the distribution functions on left and right respectively. But
such contributions  do not diverge with increasing  temperature,
which is a formal  reason for the  saturation of the odd order
correlation functions at high temperatures.

For disordered systems the transmission coefficients are random
variables and their statistical properties are known from Random
Matrix Theory (RMT) \cite{Dorokhov,Imry}. Employing RMT one may
average the scattering states result over the disorder. This
reproduces the results we have obtained in the limit
 $\l_{\rm in} \gg L$, and in particular eq.(\ref{e27}).

Finally, we briefly comment on the observability of high order
current correlation functions. Consider an ameter that detects the
net  charge transmitted through a given cross-section over a time
interval $\tau$. Under the condition that $\tau$ is sufficiently
long ($\bar{n}=\langle I\rangle\tau/e \gg 1$, where $\bar{n}$ is
the average number of  electrons passing through a given
cross-section within the time interval $\tau$), one may extract
the low frequency current fluctuations from the statistics of the
transmitted charge (known as a counting statistics). Indeed, to
describe low frequency  current fluctuations one has to require
$\hbar/ \tau $ to be smaller than the relevant energy scale. For
temperature $T \sim 100$K we require $\tau \ge 10^{-1}$p sec, the
latter condition satisfied for any practical measurements. The
more realistic restriction comes from the available electronics,
and  we estimate $\tau\sim 1$nsec. For $\langle I\rangle \sim 100$
nA we find $\bar{n} \sim 100$.
\begin{eqnarray}
\frac{\langle(n -\bar{n})^3\rangle} {(\langle(n -\bar{n})^2\rangle
)^{3/2}} \sim \frac{1}{\bar{n}^{1/2}} \approx \frac{1}{10}.
\end{eqnarray}
As we see, in principle a measurement of ${\cal{S}}_3$ is
possible. One should note that in this consideration the
nongeneric parasitic noise (such as an amplifiers $1/f$ noises and
etc.) had been ignored.

Zero temperature current noise in quantum coherent junctions (i.e.
the elastic case) \cite{Levitov_Lee}  is a manifestation of a
stochastic process that follows binomial distribution. The full
counting statistics in the elastic limit has  recently been found
\cite{Levitov_Lee,Nazarov,GGM}. It remains an open problem to find
the counting statistics in the strongly inelastic limit. Our
analysis reveals that the asymmetry of the probability
distribution function (a measure of which is ${\cal{S}}_3$) is
modified in a non-trivial way (i.e., it is either enhanced  
or suppressed by the  electron-electron 
interaction, depending on the dimensionless parameter $eV/T$).

Upon completion of this paper, we have learned of a related work
by Levitov and Reznikov (cond-mat/0111057) addressing similar
questions in quantum coherent conductors. We acknowledge
discussions with M. Reznikov, A. Kamenev, A. Mirlin  and Y. Levinson. This
research was supported
 by the U.S.-Israel Binational Science Foundation (BSF),
 by the Minerva Foundation, by the Israel Science Foundation of the Israel
 Academy of Arts and Sciences and by the German-Israel Foundation (GIF).
\appendix
\section{}
\label{appendix1} In this appendix we present the technical steps
involved in evaluation of the ${\cal{S}}_3$ in the $l_{\rm in} \ll
L$ limit. Starting  with  eq.(\ref{S3_inel}) one notes that the
part which contribute to the correlation function at question is
given by:
\begin{eqnarray}&&
\label{S3_inel_expansion_complete}
{\cal{S}}_3(t_1-t_2,t_2-t_3)=\frac{(\pi\hbar\nu
D)^2}{16}\bigg\langle
\tilde{\M}^0_1(x_1,t_1)\tilde{\I}^D_1(x_2,x_3,t_2,t_3)+
\!\frac{\pi\hbar\nu
D}{4}\!\tilde{\M}^0_1(x_1,t_1)\tilde{\M}^0_2(x_2,t_2)\tilde{\M}^0_1(x_3,t_3)\!+\nonumber
\\&&
\!\tilde{\M}^1_0(\!x_1,t_1\!)\tilde{\I}^D_1(\!x_2,x_3,t_2,t_3\!)+
\tilde{\M}^0_1(x_1,t_1)\tilde{\M}^1_1(x_2,t_2)\tilde{\M}^1_0(x_3,t_3)
+(x_1,t_1\leftrightarrow\!x_3,t_3) +
(x_1,t_1\leftrightarrow\!x_2,t_2) \bigg\rangle \,\,.
\end{eqnarray}
The first two terms entering eq.(\ref{S3_inel_expansion_complete})
are similar to those we dealt with in the non-interacting case
(cf. (\ref{z28})). The third term in
eq.(\ref{S3_inel_expansion_complete}) should be kept because out
of equilibrium the fields $\nabla k$ and $\bar{w}$ are no longer
decoupled (eq. \ref{action_out_of_equlibrium}). However, since the
coupling between those two fields is proportional to  $\nabla
\bar{F}$ we may expand the action in the latter:
\begin{eqnarray}&&
\label{S3_inel_expansion_complete2}
{\cal{S}}_3(t_1-t_2,t_2-t_3)=\frac{(\pi\hbar\nu
D)^2}{16}\bigg\langle
\tilde{\M}^0_1(x_1,t_1)\tilde{\I}^D_1(x_2,x_3,t_2,t_3)+
\!\frac{\pi\hbar\nu
D}{4}\!\tilde{\M}^0_1(x_1,t_1)\tilde{\M}^0_2(x_2,t_2)\tilde{\M}^0_1(x_3,t_3)\!+\nonumber
\\&&
\!\tilde{\M}^1_0(\!x_1,t_1\!)\tilde{\I}^D_1(\!x_2,x_3,t_2,t_3\!)iS_1^1+
\tilde{\M}^0_1(x_1,t_1)\tilde{\M}^1_1(x_2,t_2)\tilde{\M}^1_0(x_3,t_3)
+  (x_1,t_1\leftrightarrow\!x_3,t_3) +
(x_1,t_1\leftrightarrow\!x_2,t_2) \bigg\rangle_0 \,.
\end{eqnarray}
Here $\langle \rangle_0$ implies that the expectation value is
calculated employing the correlators  eqs.(\ref{c2}) and
(\ref{corr1}). The values of $\hat{\I}^D$ and $\hat{\M}$ are then
given by:
\begin{eqnarray}&&
\label{a1} \tilde{\M}^1_0(x,t)=-4\int
d[\epsilon]e^{i(\epsilon_1-\epsilon_2)t} \nabla k_{2
\epsilon_1-\epsilon_2,x}[1-\bar{F}_{x,\epsilon_1}\bar{F}_{x,\epsilon_2}]\,\,
, \\&& \tilde{\I}^0_1(x,t_2,t_3)=-2\int
d[\epsilon]e^{i(t_2(\epsilon_3-\epsilon_6)+t_3(\epsilon_5-\epsilon_4)}
[\bar{F}_{\epsilon_5,x}
w_{\epsilon_3,\epsilon_4,x}\delta_{\epsilon_5,\epsilon_6}+
\bar{F}_{\epsilon_3,x}w_{\epsilon_5,\epsilon_6,x}\delta_{\epsilon_3,\epsilon_4}]\,\,
, \\&& \tilde{\M}_1^0(x,t)\!=\!-\!2\int d[\epsilon]
e^{i(\epsilon_1-\epsilon_2)t}\! \big[
\nabla\bar{\omega}_{x,\epsilon_1,\epsilon_2}-\nabla\bar{\omega}_{x,\epsilon_1,\epsilon_2}\bar{F}_{x,\epsilon_1}\bar{F}_{x,\epsilon_2}+
\bar{\omega}_{x,\epsilon_1,\epsilon_2}\nabla
\bar{F}_{x,\epsilon_1}\bar{F}_{x,\epsilon_2}+
\bar{\omega}_{x,\epsilon_1,\epsilon_2}\bar{F}_{x,\epsilon_1}\nabla
\bar{F}_{x,\epsilon_2} \big]\,\, , \\&& \tilde{\M}_1^1(x,t)=\int
d[\epsilon]
 e^{i(\epsilon_1-\epsilon_2)t}
[\nabla {k_1}_{\epsilon_3-\epsilon_2,x}
w_{x,\epsilon_1,\epsilon_3}\!-\!\nabla
{k_1}_{\epsilon_1-\epsilon_3,x} w_{x,\epsilon_3,\epsilon_2}].
\end{eqnarray}
\begin{multicols}{2}
To evaluate  ${\cal{S}}_3(\omega_1=0,\omega_2=0)$
(eq~(\ref{S3_inel_expansion_complete})) we will use the fact that
the current fluctuations  are independent of the choice of the
cross-section; we  therefore  may integrate
eq.(\ref{S3_inel_expansion_complete}) over the entire volume of
the sample. We find that in the limit $l_{\rm in} \ll L$ \ \
${\cal{S}}_3$ has the structure (cf. eq.~(\ref{a10}))
\begin{eqnarray}&&
\label{S_3_general}
{\cal{S}}_3(\omega_1=0,\omega_2=0)=\frac{3e^3{\cal{A}}\pi g^2}{2
L^3} \int_0^L d x_1 dx_2 \int_{-\infty}^{\infty}d\epsilon_1
d\epsilon_2\nonumber \\&& F(\epsilon,x_1) \langle
w_{x,\epsilon_1,\epsilon_1}\bar{w}_{x,\epsilon_2,\epsilon_2}\rangle
\nabla\left(F^2(\epsilon,x_2)\right) \,\, .
\end{eqnarray}
Here the expectation value is calculated accordingly to the
effective action (\ref{e55}). Performing the Gaussian integration
we find the correlation function:
\begin{eqnarray}&&
\langle
w_{x,\epsilon_1,\epsilon_1}\bar{w}_{x,\epsilon_2,\epsilon_2}\rangle=
2{\cal{D}}(x_1,\epsilon_1;x_2,\epsilon_2)\,\,.
\end{eqnarray}
Here we define the zero frequency a propagator:
\begin{eqnarray}&&
\label{definition_inel_dif}
\hspace{-0.5cm}[-D\nabla^2\!+\hat{{\cal{I}}}^{ee}]{\cal{D}}(x_1,\epsilon_1;x_2,\epsilon_2)=\frac{\delta(\epsilon_1,\epsilon_2)}{\pi\nu}
\delta(x_1-x_2).
\end{eqnarray}
The propagator ${\cal{D}}$ is nothing else, but the kernel (taken
at zero frequency) of the kinetic equation (in the diffusion
approximation). In the limit when electron scattering rate is
small, the inelastic diffusion propagator becomes an ordinary
diffusion. In the opposite limit, one can invoke the perturbation
theory in $L/l^{in}$ to find its explicit form. Together with
eq.(\ref{S_3_general}) its leads us to eq.(\ref{a10_inelastic}).

\end{multicols}

\begin{thebibliography}{99}
\bibitem{experiments}
M. Reznikov, M. Heiblum, H. Shtrikman and D. Mahalu, {Phys. Rev.
Lett.} {\bf 75}, 3340 (1995); A. H. Steinbach  J. M. Martinis and
M. H. Devoret {Phys. Rev. Lett.} {\bf 76}. 3806 (1996); R.J.
Schoelkopf, P.J. Burke, A.A.Kozhevnikov, D.E.Prober and M.J.Rooks,
{Phys.Rev. Lett.} {\bf 78}, 3370 (1997); M. Henny, S. Oberholzer,
C. Strunk, and C. Schönenberger {Phys. Rev. B } {\bf 59}, 2871
(1999).


\bibitem{Kogan_book}
Sh. Kogan {\em Electronic  Noise and Fluctuations in Solids},
(Cambridge Press, 1996).

\bibitem{Blanter}
Ya. M. Blanter, M. B\"uttiker {Physics Reports} {\bf 336}, 2,
(2000).

\bibitem{de-Piccioto}
R. de-Piccioto, M. Reznikov, M. Heiblum, V. Umansky, G. Bunin and
D. Mahalu Nature {\bf 389}, 6647 (1997); L. Saminadayar, D.C.
Glattli, Y. Jin and B. Etienne Phys. Rev. Lett. {\bf 79}, 2526
(1997);

\bibitem{ALY} B.L.  Altshuler, L. S. Levitov and  A. Yu. Yakovets {Sov. Phys. JETP Lett.} {\bf 59}, 12  (1994).

\bibitem{FDT_resembles}
One may note that although the system is out of equilibrium, the
noise spectral function can be cast in terms of the equilibrium
noise function. This situation it generally true, for the
non-interacting electrons and for some special cases of the
interacting problems [E.V. Sukhorukov,G. Burkard, D. Loss Phys.
Rev. B {\bf 63}, 125315 (2001)]. In the non-interacting case the
electrons distribution function is a weighted combination of the
Fermi-Dirac distribution functions. This is in contradistinction
to the inelastic limit, where electron-electron collisions  modify
the distribution function  substantially,  cf. eq.~(\ref{cc1}).
\bibitem{Gavish}
 We note that physical observables even for the second current correlator
may be associated with  different current correlation functions,
such as symmetrised, antisymmetrzied or their combination. At low
frequency limit which we will concentrate on only symmetrised part
of survives.  For the discussion of finite frequency noise
measurements see G.B. Lesovik and R. Loosen, Pism'ma Zh. E'ksp.
Teor. Fiz. {\bf 65}, 280 (1997) [JETP Lett. {\bf 65}, 295,
(1997)]; U. Gavish, Y. Levinson and Y. Imry Phys. Rev. B. {\bf
62}, R10637 (2000).


\bibitem{KA} A. Kamenev, A. Andreev {Phys. Rev. B } {\bf 60}, 2218 (1999).

\bibitem{Nayak} C. Chamon, A.W. Ludwig and  C. Nayak {\em Phys. Rev. B} {\bf 60}, 2239 (1999);
M. V. Feigel'man, A. I. Larkin, M. A. Skvortsov Phys. Rev. B {\bf
61}, 12361 (2000)

\bibitem{GG} D.B. Gutman and Y. Gefen Phys. Rev. B. {\bf 64}, 205317 (2001).

\bibitem{Shulman} A.Ya. Shulman Sh.M. Kogan {Sov. Phys. JETP } {\bf 29} , 3
(1969). S.V. Gantsevich, V.L. Gurevich and R. Katilius {Sov. Phys.
JETP } {\bf 30}, 276 (1970).

\bibitem{Quantum_Galvanometr}L.S. Levitov, H. Lee  and G.B. Lesovik {Journal of Math. Phys.} {\bf 37}, 4845 (1996).

\bibitem{Levitov_Lesovik} L.S. Levitov and G.B. Lesovik Sov. Phys. JETP Lett.
{\bf 58}, 230, (1993).

\bibitem{Levitov_Lee} H. Lee, L.S. Levitov and A.Yu. Yakovets
{Phys. Rev. B} {\bf 51}, 4079, (1995).



\bibitem{Nagaev95} K.E. Nagaev {Phys. Rev. B} {\bf 52}, 4740, (1995).

\bibitem{Kozub}V.I. Kozub and A.M. Rudin {Phys. Rev. B}, {\bf 52}, 7853,
(1995).


\bibitem{quantum_kinetic_equation}
A. Schmid Z. Physik {\bf 271} 251 (1974); B.L. Altshuler and A.G.
Aronov JETP Lett. {\bf 30} 483 (1979).
\bibitem{AA}
B.L. Altshuler and A.G. Aronov {in \em Electron Electron
interaction in the disordered metals}, (Elsevier Science
Publishers B.V., New-York, 1985).

\bibitem{Fin}A.M. Finkel'stein Physica B {\bf 197}, 636 (1994).

\bibitem{GGM} D.B. Gutman, Y. Gefen and A. Mirlin
to appear in proceedings of "Quantum Noise", edited by Yu. V.
Nazarov and Ya. M. Blanter (Kluwer) .



\bibitem{LL} L.D. Landau and E.M. Lifshitz, {Statistical
Physics 1}, (Pergamon Press, Oxford 1980), Ch. 118 and in
particular eq.(118.7).

\bibitem{Lax} M. Lax Rev. Mod. Phys. {\bf 32}, 25 (1960).

\bibitem{deJong_Beenakker} J.M. de Jong and C.W.J. Beenakker
{\em Phys. Rev. B} {\bf 51}, 16867 (1995).

\bibitem{Nagaev2} K.E. Nagaev {Phys. Rev. B} {\bf  57}, 4628 (1998).

\bibitem{Dorokhov} O.N.  Dorokhov {Solid State Commun.} {\bf 51}, 381, (1984).

\bibitem{Imry} Y. Imry, Europhys. Lett. {\bf 1}, 249 (1986).

\bibitem{Nazarov} Yu. V. Nazarov {Ann. Phys.} (Leipzig) {\bf 8}, Spec. Issue, 193 (1999).


\bibitem{Landauer} R. Landauer, IBM J. Res. Dev. {\bf 1}, 223 (1957);
R. Landauer in {\em Localization, Interaction, and Transport
Phenomena}, (Springer-Verlag, New-York, 1984).

\end{thebibliography}
\end{document}